\documentclass[twoside,12pt]{article}

\usepackage[T1]{fontenc}
\usepackage[utf8]{inputenc}
\usepackage{indentfirst}
\usepackage{bm}
\usepackage{graphicx}
\usepackage{amsmath}
\usepackage{amssymb}

\begin{document}

\thispagestyle{empty} \vspace*{0.8cm}\hbox
to\textwidth{\vbox{\hfill\huge\sf    \hfill}}
\par\noindent\rule[3mm]{\textwidth}{0.2pt}\hspace*{-\textwidth}\noindent
\rule[2.5mm]{\textwidth}{0.2pt}


\begin{center}
\LARGE\bf Systematic Study of the \(Q\)-Value-Dependent Surface Diffuseness and \(\alpha\)-Decay Half-Lives of Superheavy Nuclei within the Coulomb and Proximity Potential Model$^{*}$
\end{center}

\footnotetext{\hspace*{-.45cm}\footnotesize $^*$ This work is supported by Yunnan Fundamental Research Projects (Grants No. 202501AT070067, No. 202401AU070074, and No. 202501CF070189), Yunnan Provincial Xing Dian Talent Support Program (Young Talents
Special Program, Grant No. XDYC-QNRC-2023-0162), Kunming University Talent Introduction Research Project (Grant No. YJL24019),  the Program for Frontier Research Team of Kunming University 2023, National Natural Science Foundation of China (Grant No. 12063006), National College Student Innovation and Entrepreneurship Training Program (Grants No. 202511393011, No. 202511393012, and No. 202511393015), Yunnan Province College Student Innovation and Entrepreneurship Training Program (Grants No. S202511393003, No. S202511393043, and No. S202511393044), and Xing Dingyu Academician Workstation of Yunnan Province (Grant No. 202605AF350035).}
\footnotetext{\hspace*{-.65cm}\footnotesize $^\dag$Corresponding author, E-mail: ygmanan@kmu.edu.cn \\  $^\ddag$Corresponding author, E-mail: yanghaitao205@163.com \\ $^{\S}$Corresponding author, E-mail: 20230001@ztu.edu.cn \\  $^{\P}$Corresponding author, E-mail: yanbingcai@mail.gufe.edu.cn}

\begin{center}
\rm Zhengwei Duan$^{\rm a)}$, \ \ Jianli Bi$^{\rm a)}$, \ Panpan Qi$^{\rm a)}$,   \ Anqi Yang$^{\rm a)}$,  \ Gongming Yu$^{\rm a)\dagger}$ \\
 \ Haitao Yang$^{\rm b)\ddagger}$,  \ Zhangyan Li$^{\rm b)\S}$, \ and  \ Yanbin Cai$^{\rm c)\P}$,
\end{center}


\begin{center}
\begin{footnotesize} \sl
${}^{\rm a)}$ College of Physics and Technology, Kunming University, Kunming 650214, China \\
${}^{\rm b)}$ School of Physics and Information Engineering, Zhaotong University, Zhaotong 657000, China \\
${}^{\rm c)}$ Guizhou Key Laboratory in Physics and Related Areas, Guizhou University of Finance and Economics, Guiyang 550025, China 
\end{footnotesize}
\end{center}



\vspace*{2mm}

\begin{center}
\begin{minipage}{15.5cm}
\parindent 20pt\footnotesize

The influence of a \(Q\)-value-dependent surface diffuseness is systematically investigated within the Coulomb and proximity potential model (CPPM) for 498 heavy and superheavy nuclei with \(Z=61\)–118. The results reveal an approximately linear dependence of the effective diffuseness on \(Q_\alpha^{1/2}\) along individual isotopic chains, with a correlation coefficient of 0.968 for nuclei with \(Z\geq100\). The calculated diffuseness values agree well with those extracted from the deformed Woods–Saxon potential, yielding a root-mean-square (RMS) deviation of 0.030 fm. The calculated \(\alpha\)-decay half-lives reproduce experimental data over 34 orders of magnitude, with an RMS logarithmic deviation of 0.277 and a correlation coefficient of 0.9992. Further analysis of the coupling between surface diffuseness and assault frequency indicates that the diffuseness primarily modifies the tunneling action, whereas nuclear structure effects are predominantly incorporated into the effective preformation factor. A reduced-mass-dependent parametrization is also examined and shows good agreement with the \(Q\)-dependent prescription for \(Z\geq104\). These findings provide a physically motivated basis for extrapolating \(\alpha\)-decay half-lives to experimentally unexplored superheavy nuclei.

\end{minipage}
\end{center}

\begin{center}
\begin{minipage}{15.5cm}
\begin{minipage}[t]{2.3cm}{\bf Keywords:}\end{minipage}
\begin{minipage}[t]{13.1cm}
Surface diffuseness; Proximity potential; Half-life; $\alpha$ decay; Superheavy nuclei 
\end{minipage}\par\vglue8pt

\end{minipage}
\end{center}

\section{Introduction}

Radioactive \(\alpha\) decay is one of the most extensively studied decay modes of heavy and superheavy nuclei and remains an indispensable experimental tool for identifying newly synthesized elements and probing nuclear shell structure near the predicted island of stability \(^{[1–4]}\). In 1928, Gamow \(^{[5]}\) and, independently, Gurney and Condon \(^{[6]}\) successfully interpreted \(\alpha\) decay in terms of quantum tunneling through the Coulomb barrier. This pioneering achievement marked one of the earliest successful applications of quantum mechanics to nuclear physics and provided a theoretical foundation for the empirical Geiger–Nuttall law \(^{[7–13]}\). Since then, a wide range of theoretical models and phenomenological approaches have been developed to describe \(\alpha\)-decay properties, including the cluster model \(^{[14–17]}\), the generalized liquid-drop model (GLDM) \(^{[18–21]}\), the universal decay law (UDL) \(^{[22,23]}\), and various microscopic approaches incorporating density-dependent interactions \(^{[24–27]}\).

Although these models differ in their treatment of \(\alpha\)-cluster preformation and assault frequency, they generally rely on a nucleus–nucleus interaction potential. While the repulsive Coulomb interaction can be reliably determined from the nuclear charge distributions, the short-range attractive nuclear interaction remains a major source of theoretical uncertainty. Different parametrizations of the nuclear potential, such as the Woods–Saxon (WS) potential \(^{[28]}\), the proximity potential derived from the liquid-drop picture \(^{[29,30]}\), and the M3Y double-folding potential \(^{[31–33]}\), generally lead to different predictions for the shape of the potential barrier. A key geometrical quantity underlying these potentials is the nuclear surface diffuseness, characterized by the diffuseness parameter \(a\) in the WS potential and the surface width \(b\) in the proximity formalism. This quantity characterizes the rate at which the nuclear density decreases from its saturation value in the nuclear interior toward zero at large distances. It plays a crucial role in determining the barrier profile, influencing the classical turning points, barrier height, and assault frequency, and ultimately affecting the predicted \(\alpha\)-decay half-life.

Early analyses of elastic-scattering data yielded surface diffuseness values of approximately 0.60–0.65 fm \(^{[34]}\). However, a systematic analysis of fusion excitation functions by Newton et al. revealed that substantially larger diffuseness values, ranging from 0.8 to 1.4 fm, were required within the standard WS parametrization to reproduce near-barrier fusion cross sections \(^{[35]}\). By combining the M3Y interaction with a repulsive term accounting for nuclear incompressibility, Ghodsi and Zanganeh obtained a diffuseness value of approximately 0.7 fm and attributed the discrepancy to the influence of nuclear incompressibility on the barrier tail \(^{[36]}\). Subsequent studies further indicated that the extracted diffuseness depends on the incident energy, the neutron-to-proton ratio \(N/Z\) of the fusing system, and the polarization of hot nuclear matter, with reported values spanning approximately 0.4–0.8 fm \(^{[37–39]}\). Collectively, these findings challenge the assumption of a universal, constant surface diffuseness parameter and suggest that its effective value may depend on the underlying nuclear properties and reaction dynamics.

In the context of \(\alpha\) decay, Dehghani, Alavi, and Benam investigated the influence of the surface diffuseness parameter on the \(\alpha\)-decay half-lives of superheavy nuclei using deformed WS and Coulomb potentials within the Wentzel–Kramers–Brillouin (WKB) approximation, and extracted a set of nucleus-dependent effective diffuseness values \(^{[40]}\). In 2019, Abdul-Latif and Omar Nagib proposed an empirical parametrization linear in \(Z\) and \(N\) within the generalized proximity formalism \(^{[41]}\), whereas Ni and Ren introduced a simple isospin-dependent expression, \(a=0.5+0.33I\) \(^{[24]}\). More recently, Prathapan et al. identified an approximately linear dependence of the effective diffuseness on \(Q_\alpha^{1/2}\) along the isotopic chains of superheavy nuclei and proposed the following \(Q\)-value-dependent parametrization, $a_{\mathrm{eff}}=(0.0030511Z-0.001501N)Q_\alpha^{1/2}+0.15455$. By incorporating this parametrization into the proximity formalism through the relation \(b=(\pi/\sqrt{3})a\), they achieved an improved description of the \(\alpha\)-decay half-lives of superheavy nuclei within the Coulomb and proximity potential model (CPPM) \(^{[42]}\). Babu et al. subsequently employed this parametrization to investigate 75 experimentally synthesized superheavy nuclei, achieving a standard deviation of approximately 0.62, and further extended its application to the prediction of \(\alpha\)-decay half-lives for yet-to-be-synthesized isotopes of elements with \(Z=119\) and 120 \(^{[43]}\).

The Coulomb and proximity potential model (CPPM) has been systematically developed by Santhosh and collaborators to describe \(\alpha\) decay and heavy-cluster radioactivity \(^{[44–49]}\). In its conventional formulation, however, the surface width is treated as a constant, with \(b=1.0\) fm. Previous investigations of the \(Q\)-dependent surface diffuseness have largely focused on relatively limited sets of superheavy nuclei, and the individual contributions of surface diffuseness and nucleus-dependent assault frequency (or the preformation factor) \(^{[50–52]}\) to the accuracy of half-life predictions have yet to be systematically disentangled across a broad region of the nuclear chart. In the present work, we incorporate the \(Q\)-dependent surface diffuseness into the CPPM and systematically calculate the \(\alpha\)-decay half-lives of 498 nuclei spanning the proton-number range \(Z=61\)–118. The resulting effective diffuseness values are compared with those extracted using the deformed WS potential and those predicted by the empirical parametrizations in Refs. [24,41]. Furthermore, we investigate the individual effects of surface diffuseness and assault frequency on the predicted half-lives and evaluate a simplified reduced-mass-dependent parametrization adopted in our computational framework.

The remainder of this paper is organized as follows. Section 2 presents the theoretical framework of the CPPM, the WKB formalism for calculating \(\alpha\)-decay half-lives, and the different parametrizations of surface diffuseness. Section 3 describes the data set and computational procedures. Section 4 presents and discusses the numerical results. Finally, Section 5 summarizes the main findings and conclusions.

\section{Theoretical framework}

\subsection{Coulomb and proximity potential model}

In the CPPM, the parent nucleus is treated as a two-body system composed of the daughter nucleus and the emitted \(\alpha\) cluster. The total interaction potential between the two fragments is the sum of the Coulomb, proximity, and centrifugal terms,
\begin{equation}
V(r)=V_{\mathrm{C}}(r)+V_{\mathrm{P}}(z)+V_\ell(r),
\label{eq:1}
\end{equation}
where \(r\) is the center-of-mass separation and \(z=r-C_1-C_2\) is the distance between the nearest surfaces. With the spherical-fragment radii \(R_i=1.28A_i^{1/3}-0.76+0.8A_i^{-1/3}\), the Süssmann central radii  can be written as
\begin{equation}
C_i=R_i-\frac{b^2}{R_i}.
\label{eq:2}
\end{equation}

The Coulomb potential is taken as the interaction of two uniformly charged spheres,
\begin{equation*}
V_{\mathrm{C}}(r)=\frac{Z_1Z_2e^2}{r},\quad r\geq R_1+R_2,
\end{equation*}
\begin{equation}
V_{\mathrm{C}}(r)=\frac{Z_1Z_2e^2}{2(R_1+R_2)}\left[3-\frac{r^2}{{(R_1+R_2)}^{2}}\right],\quad r<R_1+R_2,
\label{eq:3}
\end{equation}
with \(e^2=1.43996\) MeV fm. The centrifugal potential with the Langer correction is given by
\begin{equation}
V_\ell(r)=\frac{\hbar^2{(\ell+1/2)}^{2}}{2\mu r^2},
\label{eq:4}
\end{equation}
where \(\ell\) is the relative orbital angular momentum, and \(\mu=A_1A_2/(A_1+A_2)\) is the reduced mass in atomic-mass units (1 u = 931.5 MeV/\(c^2\) in the kinetic-energy term). The Langer form \(\ell+1/2\) is the correct radial zero-point energy in the WKB treatment.

The proximity attraction between two gently curved surfaces reads $^{[29]}$,
\begin{equation}
V_{\mathrm{P}}(z)=4\pi b\,\gamma\,R\,\Phi\!\left(\frac{z}{b}\right),\qquad R=\frac{C_1C_2}{C_1+C_2},
\label{eq:5}
\end{equation}
with the surface-energy coefficient
\begin{equation}
\gamma=0.9517\left[1-1.7826{\left(\frac{N-Z}{N+Z}\right)}^{2}\right]\ \text{MeV fm}^{-2},
\label{eq:6}
\end{equation}
and the universal proximity function used in the CPPM,
\begin{equation*}
\Phi(\varepsilon)=-0.5{(\varepsilon-2.54)}^{2}-0.0852{(\varepsilon-2.54)}^{3},\quad \varepsilon\leq1.2511,
\end{equation*}
\begin{equation}
\Phi(\varepsilon)=-3.437\,e^{-\varepsilon/0.75},\quad \varepsilon>1.2511.
\label{eq:7}
\end{equation}

Inside the contact point, the interaction potential is smoothly matched to the power-law overlap potential proposed by Blocki and Swiatecki. The surface width \(b\) enters Eq. (2), (5), and (7) through the nuclear geometry, the strength of the proximity potential, and the argument of the universal function, respectively. Consequently, it plays a crucial role in determining the shape of the potential barrier.

\subsection{WKB penetration probability}

The \(\alpha\)-decay constant is written as
\begin{equation}
\lambda=\nu P,
\label{eq:8}
\end{equation}
where \(\nu\) is the assault frequency, which effectively includes the \(\alpha\)-preformation probability, and \(P\) is the single-impact penetration probability. In the one-dimensional WKB approximation,
\begin{equation}
P=\exp\left[-\frac{2}{\hbar}\int_{R_{\mathrm{in}}}^{R_{\mathrm{out}}}\sqrt{2\mu\left|V(r)-Q_\alpha\right|}\,dr\right],
\label{eq:9}
\end{equation}
where the inner turning point is \(R_{\mathrm{in}}=R_1+R_2\) and the outer turning point follows analytically from \(V(r)=Q_\alpha\),
\begin{equation}
R_{\mathrm{out}}=\frac{Z_1Z_2e^2}{2Q_\alpha}+\sqrt{{\left(\frac{Z_1Z_2e^2}{2Q_\alpha}\right)}^{2}+\frac{\hbar^2{(\ell+1/2)}^{2}}{2\mu Q_\alpha}}.
\label{eq:10}
\end{equation}

The half-life is consequently given by
\begin{equation}
\begin{split}
T_{1/2}&=\frac{\ln2}{\nu P},\\
\log_{10}T_{1/2}&=\log_{10}(\ln2)-\log_{10}\nu-\frac{1}{\ln10}\ln P.
\end{split}
\label{eq:11}
\end{equation}

Following the original CPPM implementation, the assault frequency is parameterized as \(\nu=10^{P_a}\times10^{22}\ \mathrm{s}^{-1}\), where \(P_a\) is an empirical, nucleus-dependent quantity (\(\log_{10}\nu\) ranges from 18.5 to 21.1 in the present sample) that absorbs both the preformation probability and the internal kinematics. 

\subsection{Surface diffuseness prescriptions}

The standard CPPM adopts \(b=1.0\) fm, corresponding to a WS diffuseness \(a=\sqrt{3}b/\pi\approx0.55\) fm. The main prescription used here is the \(Q\)-value-dependent effective diffuseness of Prathapan \emph{et al.} $^{[42]}$,
\begin{equation}
a_{\mathrm{eff}}=(0.0030511Z-0.001501N)\,Q_\alpha^{1/2}+0.15455\ \text{fm},
\label{eq:12}
\end{equation}
and then the proximity width is given by
\begin{equation}
b_{\mathrm{eff}}=\frac{\pi}{\sqrt{3}}\,a_{\mathrm{eff}}.
\label{eq:13}
\end{equation}

Two further empirical prescriptions are evaluated for comparison: the \(Z\),\(N\)-linear formula of Abdul-Latif and Omar Nagib $^{[41]}$,
\begin{equation}
a_{\mathrm{AL}}=0.012063Z+0.0019759N-1.09535
\label{eq:14}
\end{equation}
and the isospin formula of Ni and Ren $^{[24]}$,
\begin{equation}
a_{\mathrm{NR}}=0.5+0.33I,\qquad I=\frac{N-Z}{A}.
\label{eq:15}
\end{equation}

Finally we examine a simpler single-variable parametrization, the one implemented in our own code (CPPM3),
\begin{equation}
a_{\mathrm{eff}}^{(\mu)}=12.4758\,\mu-48.727,
\label{eq:16}
\end{equation}
which is obtained by a least-squares fit to the reduced mass in the superheavy region and reflects, phenomenologically, the dependence of the surface geometry on fragment mass asymmetry. Because \(\mu\) can be small for light systems, Eq. (16) yields unphysical nonpositive values for 78 of the 498 nuclei and is therefore restricted to the heavy and superheavy region (\(Z\gtrsim82\)). The 498-nucleus main results presented below are generated with Eqs. (12)--(13), which reproduce the data file exactly; Eq. (16) is retained as an auxiliary approximation in this region.

\section{Data and numerical procedure}

The data set contains 498 \(\alpha\)-decay entries with parent proton number \(Z=61\)--118, mass number \(A=145\)--294, decay energy \(Q_\alpha=2.20\)--11.84 MeV, and experimental half-lives spanning \(\log_{10}T_{1/2}^{\mathrm{exp}}=-7.16\)--26.80. It includes 359 nuclei with \(Z\geq82\), 60 with \(Z\geq100\), and 33 with \(Z\geq104\). The \(Q\) values and experimental half-lives are taken from the NNDC/ENSDF and NUBASE2020 evaluations $^{[53,54]}$ together with the AME2016/AME2020 atomic-mass evaluations $^{[55,56]}$, consistently with earlier CPPM analyses. The data comprise 172 even--even, 142 even-\(Z\) odd-\(N\), 119 odd-\(Z\) even-\(N\), and 65 odd--odd nuclei, and contain 354 favored transitions with \(\ell=0\).

The calculation is implemented in Python 3 with NumPy and SciPy. Apart from replacing the fixed statement \texttt{b\ =\ 1.0} by the nucleus-dependent \(b_{\mathrm{eff}}\) of Eqs. (12)--(13), all ingredients---radii, proximity, Coulomb, and centrifugal potentials, turning points, and the action integral---are identical to the original CPPM code. The action integral in Eq. (9) is evaluated with the adaptive routine \texttt{scipy.integrate.quad}. An independent reproduction of the tabulated \(\log_{10}T_{1/2}\) values agrees with the output file to better than \(10^{-14}\), confirming the numerical consistency of the code and data. The predictive quality is quantified by the root-mean-square deviation
\begin{equation}
\sigma_{\mathrm{RMS}}=\sqrt{\frac{1}{n}\sum_{i=1}^{n}\left[\log_{10}T_{1/2,i}^{\mathrm{cal}}-\log_{10}T_{1/2,i}^{\mathrm{exp}}\right]^2},
\label{eq:17}
\end{equation}
and by the mean absolute deviation, using the same convention as in related works on the same 498-nucleus sample $^{[57,58]}$.

\section{Results and discussion}

\subsection{Linear dependence of the effective diffuseness on $Q_\alpha^{1/2}$}

The effective surface width \(b_{\mathrm{eff}}\) as a function of \(Q_\alpha^{1/2}\) for all 498 nuclei, with the data points color-coded according to their mass numbers and a global linear fit included for comparison is shown in Fig.1. Despite the broad proton-number range (\(Z=61\)–118) and the considerable scatter that may partly arise from nuclear shell effects, the overall trend is reasonably well described by the linear relation \(b_{\mathrm{eff}}=0.2058Q_\alpha^{1/2}+0.1354\), with a coefficient of determination of \(R^2=0.649\), corresponding to a Pearson correlation coefficient of 0.806. When the analysis is restricted to superheavy nuclei with \(Z=100\)–116, as shown in Fig. 2, distinct isotopic chains emerge, each exhibiting an approximately linear dependence of \(b_{\mathrm{eff}}\) on \(Q_\alpha^{1/2}\). This behavior is consistent with the findings of Prathapan et al. based on a smaller data set. Along a given isotopic chain, lighter isotopes generally exhibit larger values of both \(Q_\alpha\) and \(b_{\mathrm{eff}}\), whereas the negative neutron-number-dependent term, \(-0.001501N\), in Eq. (12) contributes to a reduction in the effective surface width for heavier, more neutron-rich isotopes. The strength of these linear correlations is further supported by correlation coefficients exceeding 0.9 for most isotopic chains, with representative values of 0.988 for \(Z=100\), 0.999 for \(Z=108\), and at least 0.998 for \(Z=112\), 114, and 116.

\begin{center}
\includegraphics[width=0.8\textwidth]{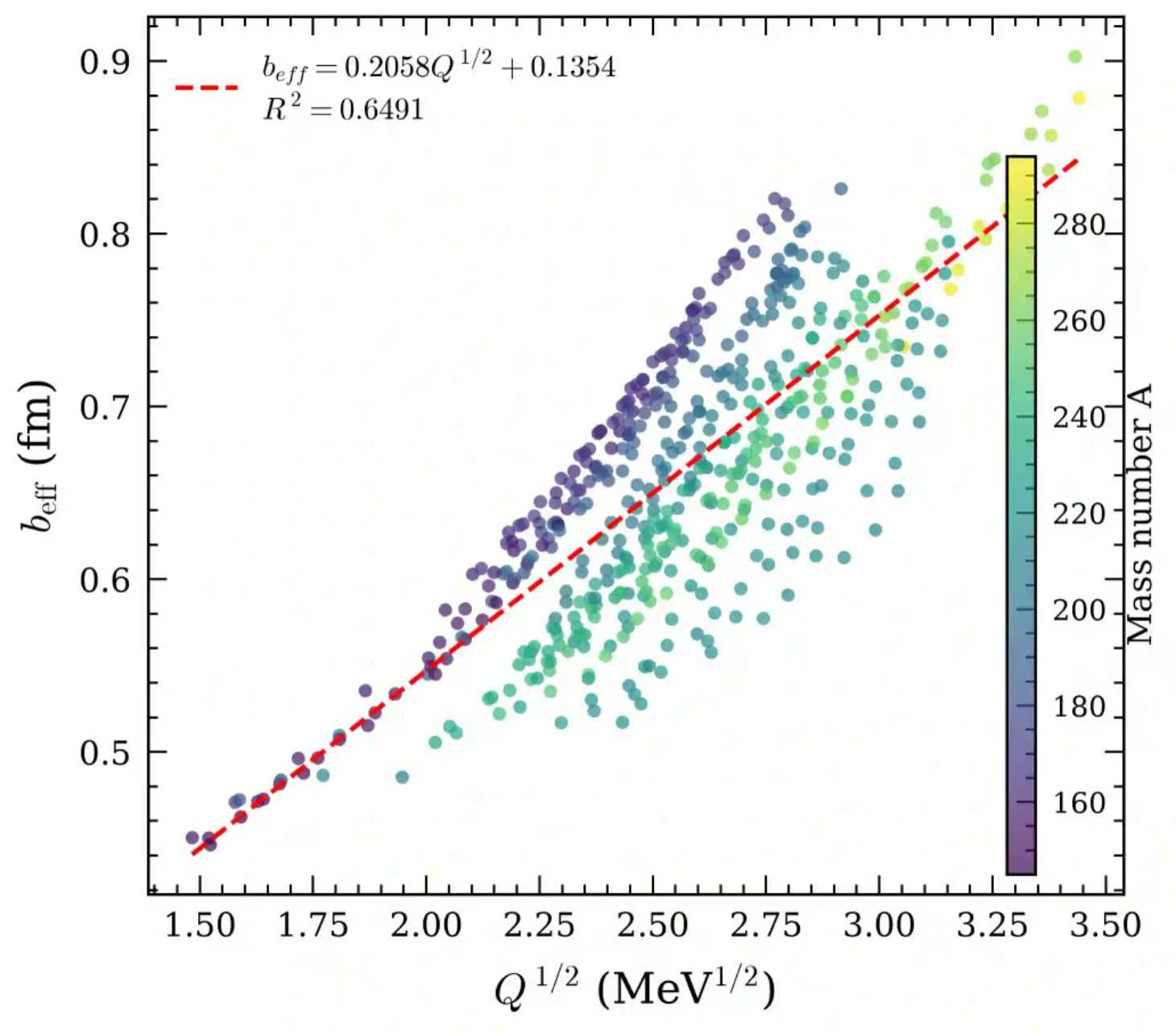}
\parbox{15.5cm}{\small{\bf Fig.1.} Surface width \(b_{\mathrm{eff}}\) as a function of \(Q_\alpha^{1/2}\) for the 498 \(\alpha\) emitters, colored by mass number. The dashed line is the global linear fit \(b_{\mathrm{eff}}=0.2058Q_\alpha^{1/2}+0.1354\) (\(R^2=0.649\)).}
\end{center}

The Pearson correlation coefficient between \(a_{\mathrm{eff}}\) and \(Q_\alpha^{1/2}\) increases from 0.806 for the full data set to 0.968 for nuclei with \(Z\geq100\) and remains relatively high at 0.933 for those with \(Z\geq104\). These results indicate that the \(Q_\alpha^{1/2}\) dependence is more pronounced in the superheavy region, whereas the greater scatter observed for lighter nuclei may reflect the combined effects of nuclear shell structure, deformation, and variations in the \(\alpha\)-particle preformation probability. Across the entire data set, \(a_{\mathrm{eff}}\) ranges from 0.246 to 0.498 fm, corresponding to \(b_{\mathrm{eff}}\) values of 0.446–0.903 fm. For nuclei with \(Z\geq104\), the effective surface width is predominantly distributed within the range of 0.69–0.90 fm, which overlaps with the larger surface diffuseness values inferred from near-barrier fusion analyses. This agreement suggests a possible connection between the effective nuclear surface properties governing \(\alpha\) decay and those influencing nuclear fusion, providing motivation for a consistent treatment of surface diffuseness in both processes.

\subsection{Comparison with the deformed Woods--Saxon benchmark}

The Table 1 summarizes the \(\alpha\)-decay energies, surface diffuseness values obtained from four different prescriptions, and the corresponding half-lives calculated within the CPPM for 33 nuclei with \(Z=104\)–118, following the format adopted by Prathapan et al. For the ten nuclei also investigated by Dehghani et al. using the deformed WS potential, the present \(a_{\mathrm{eff}}\) values exhibit excellent agreement with their results, yielding a root-mean-square (rms) deviation of only 0.030 fm, compared with the value of 0.038 fm reported by Prathapan et al. for a set of 30 nuclei. As illustrated in Fig. 3, the four diffuseness prescriptions exhibit markedly different systematic trends. The \(Q\)-dependent parametrization closely reproduces the reference values extracted using the deformed WS potential, whereas the parametrization linear in \(Z\) and \(N\) predicts a monotonic increase in surface diffuseness with increasing mass number, exceeding 0.65 fm for the heaviest nuclei considered. In contrast, the isospin-dependent parametrization yields nearly constant diffuseness values of approximately 0.56 fm, failing to capture the pronounced variation in \(a_{\mathrm{eff}}\) with \(Q_\alpha\). Both the parametrization linear in \(Z\) and \(N\) and the isospin-dependent parametrization predict a generally monotonic increase in surface diffuseness with nuclear size, a trend that is not supported by the reference values extracted using the deformed WS potential. In contrast, the explicit incorporation of \(Q_\alpha\) enables the empirical parametrization to capture variations in surface diffuseness both across different isotopic chains and along individual chains. From a physical perspective, the effective surface diffuseness should not be regarded solely as a static geometrical quantity, as it may also reflect the energetics associated with \(\alpha\)-cluster formation. The surface energy involved in this process is related to the decay energetics, analogous to the dependence of the surface energy coefficient on neutron–proton asymmetry introduced through the asymmetry term in Eq. (6). These considerations provide a physical motivation for incorporating both nuclear composition and decay energy into the parametrization of effective surface diffuseness.

\begin{center}
\includegraphics[width=0.8\textwidth]{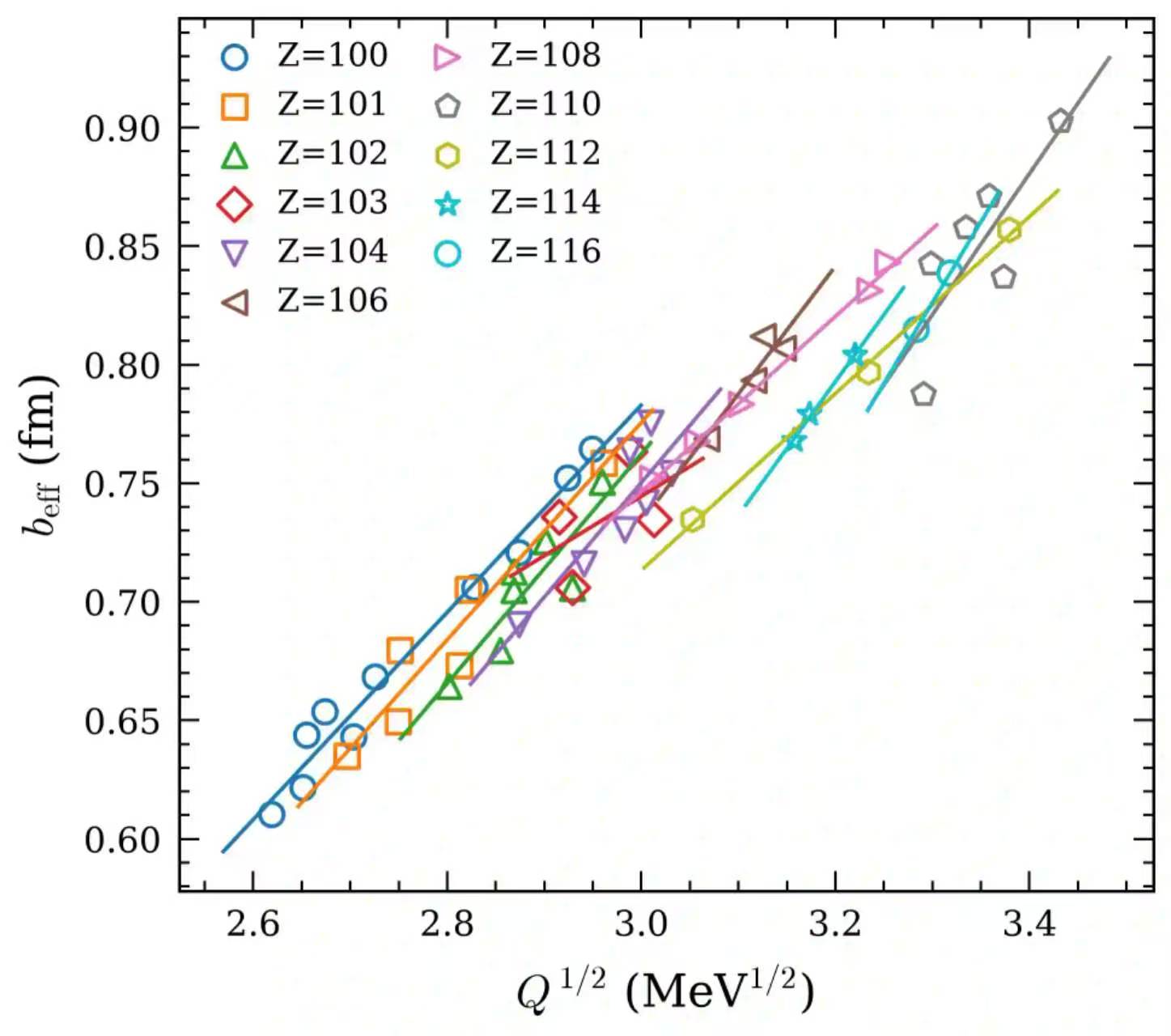}
\parbox{15.5cm}{\small{\bf Fig.2.} Surface width \(b_{\mathrm{eff}}\) versus \(Q_\alpha^{1/2}\) for the superheavy nuclei with \(Z=100\)--116. Different symbols and colors denote different proton numbers, and the solid lines are least-squares fits to the individual isotopic chains.}
\end{center}

For superheavy nuclei with \(Z\geq104\), the reduced-mass-dependent parametrization given by Eq. (16) yields effective surface widths \(b_{\mathrm{eff}}\) ranging from 0.714 to 0.902 fm, in close agreement with the range of 0.690–0.903 fm obtained from Eq. (12). The corresponding rms deviations of the calculated \(\alpha\)-decay half-lives are 0.292 and 0.296, respectively, indicating comparable predictive performance in the superheavy region. These results suggest that the reduced mass may serve as an effective proxy for the combined dependence on \(Z\), \(N\), and \(Q_\alpha\) in the parametrization of surface diffuseness for superheavy nuclei. Equation (16) therefore provides a simplified alternative for estimating the effective surface width when only nuclear mass information is available. However, its applicability to lighter nuclei is limited, as evidenced by the 78 unphysical diffuseness values obtained in the lower-\(Z\) region. Consequently, Eq. (12) is adopted for the systematic analysis of the full data set.

\begin{center}
\includegraphics[width=0.86\textwidth]{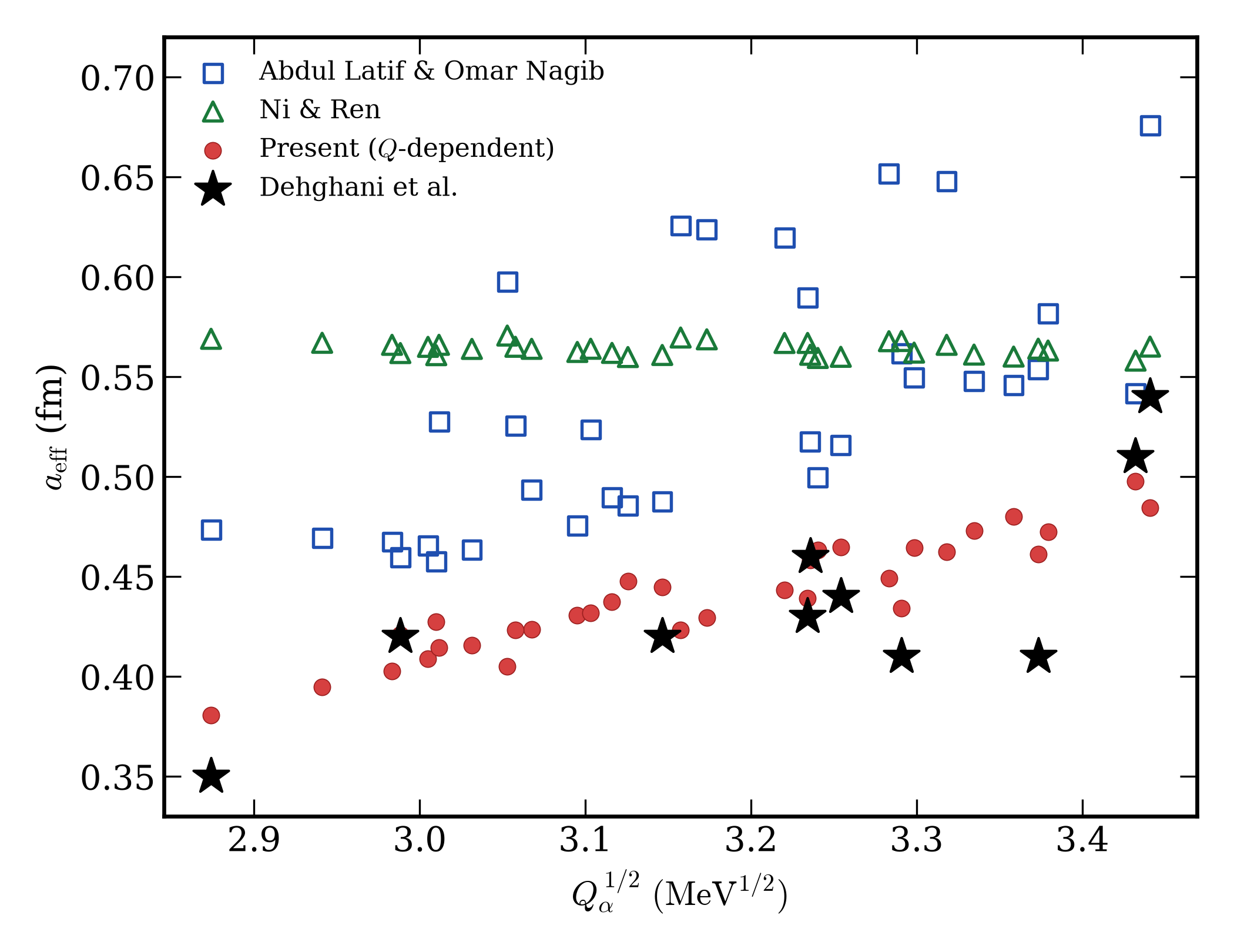}
\parbox{15.5cm}{\small{\bf Fig.3.} Effective diffuseness of the 33 superheavy nuclei as a function of \(Q_\alpha^{1/2}\): present \(Q\)-dependent results (red circles) compared with the deformed Woods--Saxon values of Dehghani \emph{et al.} (black stars), the \(Z\),\(N\)-linear formula of Abdul-Latif and Omar Nagib (open blue squares), and the isospin formula of Ni and Ren (open green triangles).}
\end{center}

\subsection{Large-sample half-life results}

The experimental \(\alpha\)-decay half-lives of superheavy nuclei with \(Z=100\)–116 as a function of the corresponding effective surface width \(b_{\mathrm{eff}}\) are shown in Fig. 4. With the increasing proton

\begin{center}
\tabcolsep=5pt
\scriptsize
\renewcommand\arraystretch{1.32}
\begin{minipage}{17cm}{\small{\bf Table 1.} The \(Q\)-dependent effective surface diffuseness and the corresponding \(\alpha\)-decay half-lives of 33 superheavy nuclei. The present \(a_{\mathrm{eff}}\) and \(b_{\mathrm{eff}}\) are compared with the values of Dehghani \emph{et al.} and with those of Abdul Latif \emph{et al.}\ and Ni \& Ren; the last two columns give the calculated and experimental \(\log_{10}T_{1/2}\).}
\end{minipage}
\vglue5pt
\begin{tabular}{lcccccccc}
\hline
\multicolumn{7}{c}{} & \multicolumn{2}{c}{\(\log_{10}T_{1/2}\)}\\
\cline{8-9}
Parent & \(Q_\alpha\) (MeV) & \(a_{\mathrm{eff}}\) & Abdul Latif & Ni \& Ren & Present \(a_{\mathrm{eff}}\) & \(b_{\mathrm{eff}}\) (fm) & cal & exp\\
\hline
$^{255}$Rf & 9.06 & --- & 0.458 & 0.561 & 0.427 & 0.775 & 1.02 & 0.54 \\
$^{256}$Rf & 8.93 & 0.42 & 0.460 & 0.562 & 0.421 & 0.764 & 0.37 & 0.32 \\
$^{258}$Rf & 9.19 & --- & 0.463 & 0.564 & 0.416 & 0.754 & $-0.55$ & $-0.98$ \\
$^{259}$Rf & 9.03 & --- & 0.465 & 0.565 & 0.409 & 0.742 & 0.39 & 0.46 \\
$^{260}$Rf & 8.90 & --- & 0.467 & 0.566 & 0.403 & 0.730 & 0.25 & 0.02 \\
$^{261}$Rf & 8.65 & --- & 0.469 & 0.567 & 0.395 & 0.716 & 1.30 & 0.90 \\
$^{263}$Rf & 8.26 & 0.35 & 0.473 & 0.569 & 0.381 & 0.690 & 3.65 & 3.34 \\
$^{259}$Db & 9.58 & --- & 0.476 & 0.562 & 0.431 & 0.781 & $-0.24$ & $-0.29$ \\
$^{259}$Sg & 9.77 & --- & 0.486 & 0.560 & 0.448 & 0.812 & $-0.35$ & $-0.38$ \\
$^{260}$Sg & 9.90 & 0.42 & 0.488 & 0.561 & 0.445 & 0.807 & $-1.77$ & $-1.91$ \\
$^{261}$Sg & 9.71 & --- & 0.490 & 0.562 & 0.437 & 0.793 & $-0.43$ & $-0.73$ \\
$^{263}$Sg & 9.41 & --- & 0.494 & 0.564 & 0.424 & 0.769 & $-0.14$ & 0.03 \\
$^{261}$Bh & 10.50 & --- & 0.500 & 0.559 & 0.463 & 0.841 & $-1.87$ & $-1.87$ \\
$^{264}$Hs & 10.59 & 0.44 & 0.516 & 0.560 & 0.465 & 0.843 & $-2.88$ & $-2.97$ \\
$^{265}$Hs & 10.47 & 0.46 & 0.518 & 0.561 & 0.458 & 0.831 & $-2.35$ & $-2.71$ \\
$^{268}$Hs & 9.63 & --- & 0.524 & 0.564 & 0.432 & 0.783 & $-0.29$ & 0.15 \\
$^{269}$Hs & 9.35 & --- & 0.526 & 0.565 & 0.423 & 0.768 & 0.97 & 1.20 \\
$^{270}$Hs & 9.07 & --- & 0.528 & 0.566 & 0.415 & 0.752 & 1.27 & 0.95 \\
$^{267}$Ds & 11.78 & 0.51 & 0.542 & 0.558 & 0.498 & 0.903 & $-4.80$ & $-5.00$ \\
$^{269}$Ds & 11.28 & --- & 0.546 & 0.560 & 0.480 & 0.871 & $-3.68$ & $-3.64$ \\
$^{270}$Ds & 11.12 & --- & 0.548 & 0.561 & 0.473 & 0.858 & $-3.54$ & $-3.69$ \\
$^{271}$Ds & 10.88 & --- & 0.550 & 0.562 & 0.464 & 0.842 & $-1.70$ & $-1.05$ \\
$^{273}$Ds & 11.38 & 0.41 & 0.554 & 0.564 & 0.461 & 0.837 & $-3.47$ & $-3.62$ \\
$^{277}$Ds & 10.83 & 0.41 & 0.562 & 0.568 & 0.434 & 0.787 & $-1.99$ & $-2.22$ \\
$^{277}$Cn & 11.42 & --- & 0.582 & 0.563 & 0.472 & 0.857 & $-3.34$ & $-3.07$ \\
$^{281}$Cn & 10.46 & 0.43 & 0.590 & 0.567 & 0.439 & 0.797 & $-0.57$ & $-0.74$ \\
$^{285}$Cn & 9.32 & --- & 0.598 & 0.571 & 0.405 & 0.735 & 2.24 & 1.51 \\
\hline
\end{tabular}
\end{center}

\begin{center}
\tabcolsep=5pt
\scriptsize
\renewcommand\arraystretch{1.32}
\begin{minipage}{17cm}{\small{\bf Table 1 (Continued).} The \(Q\)-dependent effective surface diffuseness and the corresponding \(\alpha\)-decay half-lives of 33 superheavy nuclei. The present \(a_{\mathrm{eff}}\) and \(b_{\mathrm{eff}}\) are compared with the values of Dehghani \emph{et al.} and with those of Abdul Latif \emph{et al.}\ and Ni \& Ren; the last two columns give the calculated and experimental \(\log_{10}T_{1/2}\).}
\end{minipage}
\vglue5pt
\begin{tabular}{lcccccccc}
\hline
\multicolumn{7}{c}{} & \multicolumn{2}{c}{\(\log_{10}T_{1/2}\)}\\
\cline{8-9}
Parent & \(Q_\alpha\) (MeV) & \(a_{\mathrm{eff}}\) & Abdul Latif & Ni \& Ren & Present \(a_{\mathrm{eff}}\) & \(b_{\mathrm{eff}}\) (fm) & cal & exp\\
\hline
$^{286}$Fl & 10.37 & --- & 0.620 & 0.567 & 0.443 & 0.804 & $-0.63$ & $-0.46$ \\
$^{288}$Fl & 10.07 & --- & 0.624 & 0.569 & 0.430 & 0.779 & 0.03 & $-0.12$ \\
$^{289}$Fl & 9.97 & --- & 0.626 & 0.570 & 0.423 & 0.768 & 0.55 & 0.38 \\
$^{290}$Lv & 11.01 & --- & 0.648 & 0.566 & 0.462 & 0.839 & $-1.81$ & $-2.10$ \\
$^{292}$Lv & 10.78 & --- & 0.652 & 0.568 & 0.449 & 0.815 & $-1.27$ & $-1.62$ \\
$^{294}$Og & 11.84 & 0.54 & 0.676 & 0.565 & 0.484 & 0.879 & $-3.11$ & $-2.94$ \\
\hline
\end{tabular}
\end{center}

\noindent number, \(b_{\mathrm{eff}}\) increases from approximately 0.60 to 0.92 fm, while the half-lives decrease by more than ten orders of magnitude, primarily reflecting the strong dependence of \(\alpha\)-decay half-lives on \(Q_\alpha\). Within individual isotopic chains, however, the relatively small variation in \(b_{\mathrm{eff}}\) makes it difficult to disentangle its contribution from that of the decay energy. For the full data set, the calculated half-lives exhibit excellent agreement with the experimental values over approximately 34 orders of magnitude, spanning from \(\log_{10}(T_{1/2}/\mathrm{s})\approx-7\) to approximately 27. Calculations incorporating the \(Q\)-dependent surface diffuseness yield a Pearson correlation coefficient of \(R=0.9992\), an rms logarithmic deviation of \(\sigma_{\mathrm{RMS}}=0.277\), and a mean absolute logarithmic deviation of 0.19. Only one nucleus exhibits a deviation exceeding one order of magnitude, and none exceeds two orders of magnitude. The rms logarithmic deviation of 0.277 corresponds to a characteristic multiplicative deviation of approximately 1.89, indicating a high level of agreement between the calculated and experimental half-lives. This predictive accuracy is comparable to that reported for contemporary theoretical and computational approaches, including the unified fission model, improved GLDM, and machine-learning-based methods applied to the same data set.

The comparison of the calculated and experimental \(\alpha\)-decay half-lives for all 498 nuclei using the two diffuseness prescriptions are shown in Fig. 5. Both sets of results closely follow the diagonal over the entire range of experimental half-lives. However, the results obtained with the \(Q\)-dependent surface diffuseness (red circles) are systematically shifted upward by approximately 0.17 dex relative to those obtained with the fixed-\(b\) prescription (open blue downward-pointing triangles). This systematic shift reflects an increase in the WKB tunneling action arising from the modification of the potential barrier by the \(Q\)-dependent surface diffuseness.

\begin{center}
\includegraphics[width=0.8\textwidth]{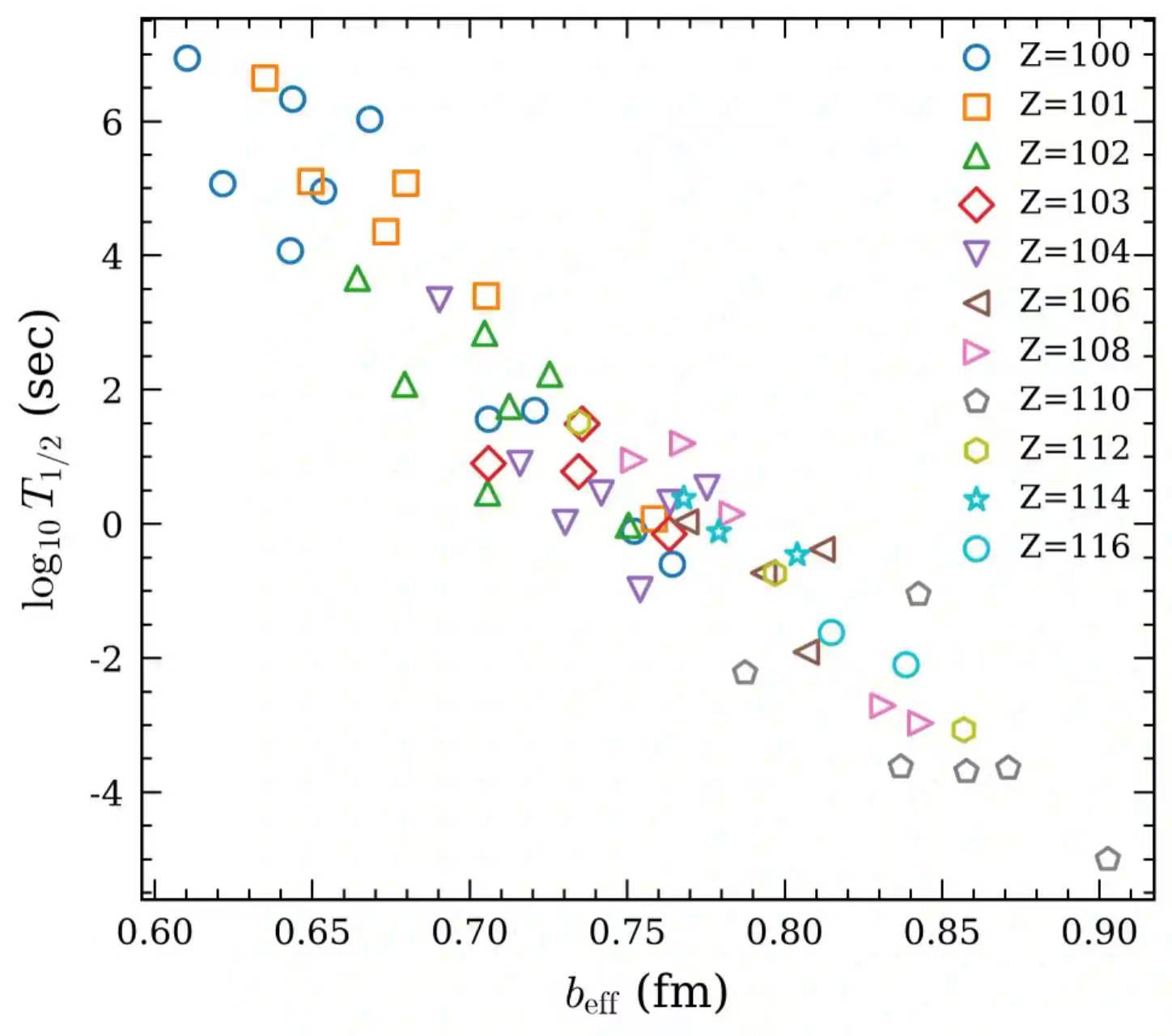}
\parbox{15.5cm}{\small{\bf Fig.4.} Experimental \(\log_{10}T_{1/2}\) of the superheavy nuclei with \(Z=100\)--116 as a function of \(b_{\mathrm{eff}}\). Different symbols and colors denote different proton numbers.}
\end{center}

The Fig. 6 presents the logarithmic deviations between the calculated and experimental \(\alpha\)-decay half-lives as a function of the parent nuclear mass number \(A\). The results obtained with the \(Q\)-dependent surface diffuseness are predominantly confined within the \(\pm0.5\) band across the entire mass range. In contrast, the fixed-\(b\) results exhibit a pronounced mass-dependent trend, with deviations generally below \(-0.2\) for \(A\gtrsim220\) and close to zero for lighter nuclei. The residual distribution is approximately Gaussian, with a mean value of \(+0.17\), indicating a slight systematic overestimation. Table 2 summarizes the rms deviations for different nuclear types. The best agreement with experimental data is achieved for even–even nuclei (\(\sigma_{\mathrm{RMS}}=0.244\)), whereas nuclei containing unpaired nucleons exhibit somewhat larger deviations. This behavior may be attributed, in part, to the blocking effect of unpaired nucleons on \(\alpha\)-cluster formation in odd-\(A\) and odd–odd nuclei. Further classification according to the orbital angular momentum carried by the emitted \(\alpha\) particle yields rms deviations of 0.254 for favored transitions with \(\ell=0\) and 0.359 for hindered transitions with \(\ell=2\). The relatively larger deviations for hindered transitions may reflect the additional influence of the centrifugal barrier and angular-momentum-dependent nuclear structure effects.

\begin{center}
\includegraphics[width=0.72\textwidth]{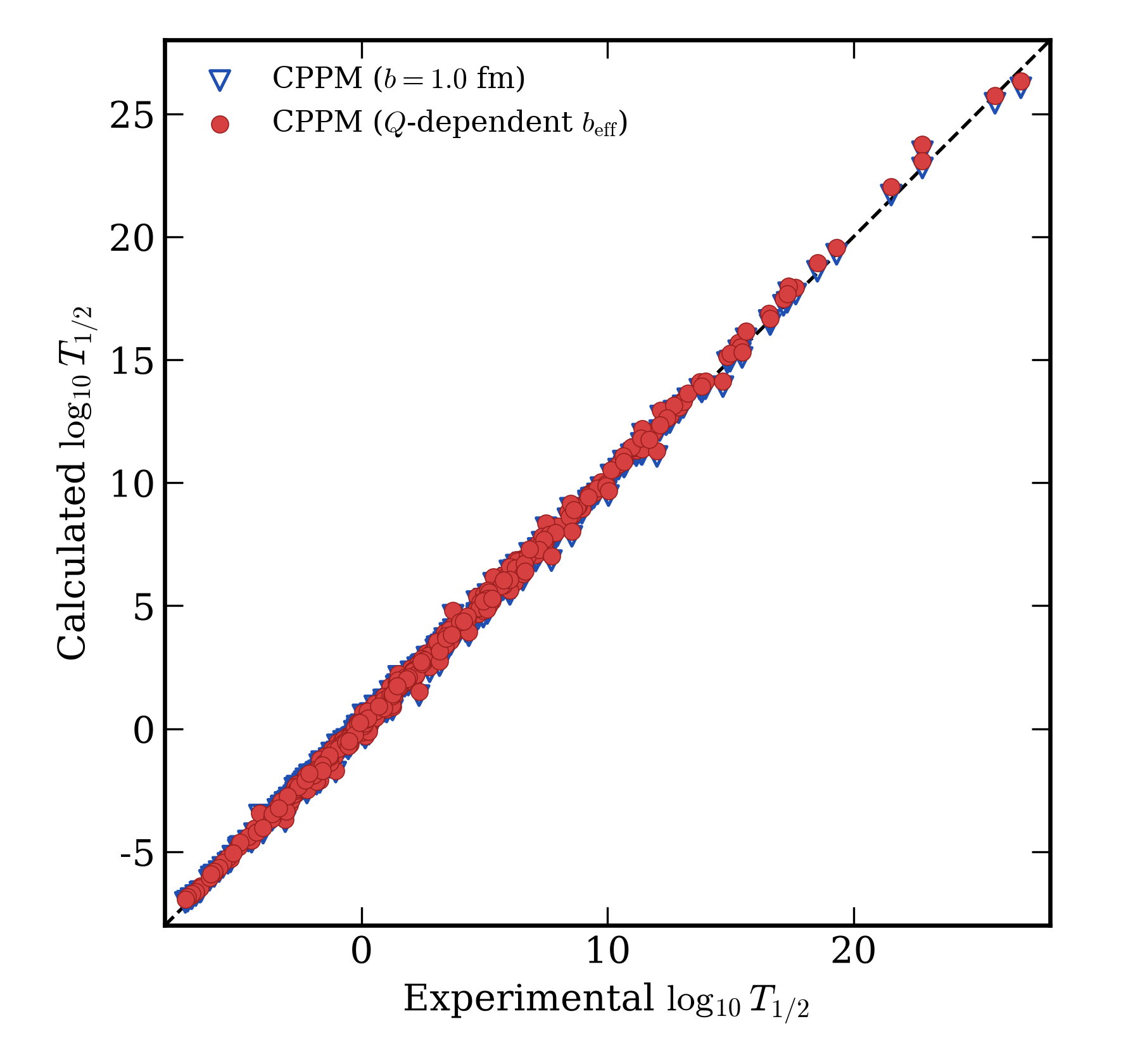}
\parbox{15.5cm}{\small{\bf Fig.5.} Calculated versus experimental \(\log_{10}T_{1/2}\) for the 498 \(\alpha\) emitters. Open blue down-triangles denote the standard CPPM with \(b=1.0\) fm; red circles denote the \(Q\)-dependent diffuseness. The dashed line represents perfect agreement.}
\end{center}

\subsection{Coupling of diffuseness and assault frequency}

One result presented in Table 2 warrants particular attention. When the empirical nucleus-dependent assault frequency is retained, the calculation with a constant surface width of \(b=1.0\) fm yields a slightly smaller rms deviation (0.212) than that obtained with the \(Q\)-dependent surface diffuseness (0.277). However, this result should not be interpreted as evidence in favor of the constant-width approximation. As indicated by Eqs. (8) and (11), the \(\alpha\)-decay half-life is jointly determined by the barrier penetrability \(P\) and the assault frequency \(\nu\). The empirical \(P_a\) values were individually extracted for all 498 nuclei under the assumption of a constant surface width of \(b=1.0\) fm. Consequently, any systematic bias in the barrier penetrability arising from this assumption may already have been absorbed into the empirical assault frequency. Modifying the surface width while retaining the previously extracted assault frequencies therefore introduces an inconsistency between the two components of the decay model. Under these circumstances, a direct comparison of the rms deviations does not provide an independent assessment of the physical validity of the respective diffuseness prescriptions.

To ensure a consistent comparison among the different diffuseness prescriptions, the assault frequency was fixed at \(\nu=10^{20}\ \mathrm{s}^{-1}\), and a single global normalization factor was introduced for each prescription, effectively allowing the adjustment of one overall prefactor. Within this framework, the rms deviations obtained using \(b=1.00\) fm, the scattering-derived value of \(b=0.63\) fm, and the \(Q\)-dependent \(b_{\mathrm{eff}}\) differ by less than 0.02 for both the full data set and the subsets with \(Z\geq82\) and \(Z\geq104\). The reduced-mass-dependent parametrization yields marginally better agreement in the superheavy region. These results indicate that, within the spherical one-dimensional WKB framework, variations in surface diffuseness primarily introduce a systematic shift in the tunneling action, with relatively little influence on the shape of the residual distribution. Once an adjustable global prefactor is introduced, modifying the spherical surface diffuseness alone provides limited scope for further reducing the rms deviation across a large nuclear data set. This observation is consistent with the adoption of parity-dependent preformation factors, deformed nuclear potentials, and microscopic double-folding potentials in more sophisticated theoretical approaches aimed at improving the accuracy of \(\alpha\)-decay half-life predictions.

\begin{center}
\includegraphics[width=0.86\textwidth]{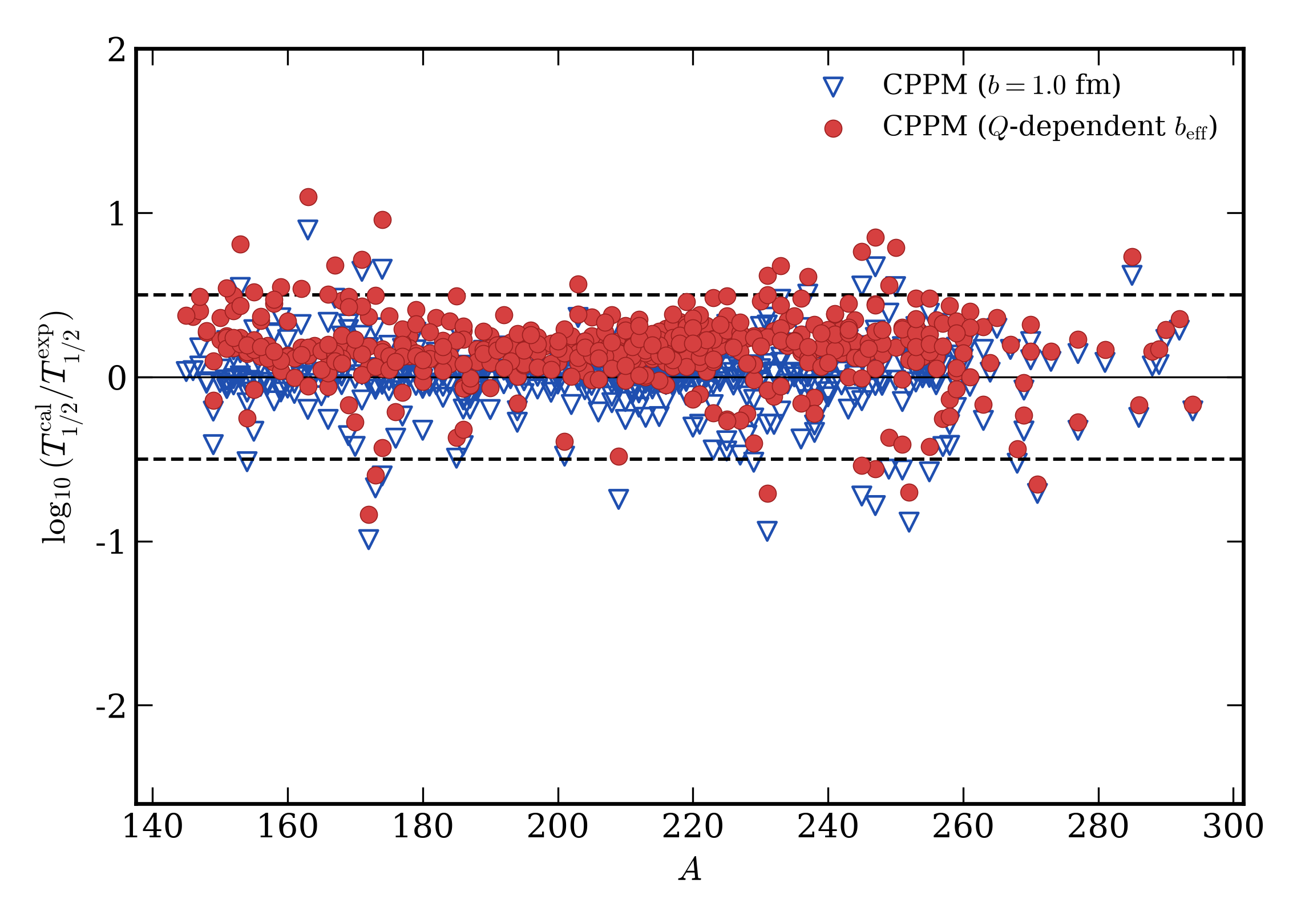}
\parbox{15.5cm}{\small{\bf Fig.6.} Logarithmic deviations between calculated and experimental \(\alpha\)-decay half-lives as a function of the mass number \(A\). The dashed lines mark \(\pm0.5\) and the solid line marks zero.}
\end{center}

\begin{center}
\tabcolsep=8pt
\small
\renewcommand\arraystretch{1.2}
\begin{minipage}{15.5cm}{\small{\bf Table 2.} Root-mean-square deviations $\sigma_{\mathrm{RMS}}$ of the two CPPM calculations for different nuclear types.}
\end{minipage}
\vglue5pt
\begin{tabular}{lcccc}
\hline
Nuclear type & \(n\) & CPPM, \(b=1.0\) fm & CPPM, \(Q\)-dependent \(b_{\mathrm{eff}}\)\\
\hline
Even--even & 172 & 0.144 & 0.244\\
Even--odd & 142 & 0.238 & 0.300\\
Odd--even & 119 & 0.236 & 0.299\\
Odd--odd & 65 & 0.249 & 0.268\\
All nuclei & 498 & 0.212 & 0.277\\
\hline
\end{tabular}
\end{center}

The essential distinction among these diffuseness prescriptions therefore lies not primarily in their numerical accuracy but in their physical basis and predictive applicability. The favorable performance of the conventional prescription with \(b=1.0\) fm under the empirical-frequency treatment partly reflects the calibration of nucleus-dependent assault frequencies against experimental half-lives. In contrast, the \(Q\)-dependent effective surface diffuseness is determined from a previously established parametrization motivated by deformed WS calculations and studies of nuclear fusion, without requiring additional nucleus-specific adjustments to reproduce the half-life under consideration. It therefore provides a physically motivated input for extrapolations to experimentally unexplored nuclei, as demonstrated by the applications of Babu et al. To distinguish the contributions of surface diffuseness from those associated with \(\alpha\)-particle preformation, future comparative studies should report results obtained under both treatments: empirical nucleus-dependent assault frequencies and a unified assault frequency supplemented by a single global normalization factor. Such a comparison would facilitate a more transparent assessment of the respective roles of surface diffuseness and preformation effects in determining \(\alpha\)-decay half-lives.

\subsection{Relation to semi-empirical half-life formulas}

Over the past several decades, numerous analytical and semi-empirical formulas have been developed to describe \(\alpha\)-decay half-lives, including the Viola–Seaborg–Sobiczewski (VSS) formula, the Brown formula, the analytical formula of Royer, the universal curve of Poenaru et al., and the universal decay law (UDL) of Qi et al. These formulas incorporate the principal dependence of the tunneling exponent on nuclear charge and decay energy and generally achieve satisfactory accuracy within their respective calibration regions. However, their reliability when extrapolated to nuclear regions far beyond the available experimental data remains uncertain. In contrast, the CPPM is formulated on the basis of nuclear geometry and the liquid-drop description of nuclear interactions. Incorporating a physically motivated surface diffuseness into this framework directly modifies the radial profile of the interaction potential, rather than merely adjusting the coefficients of an analytical half-life formula. The rms deviation of 0.277 obtained in the present work is comparable to, or smaller than, the values of approximately 0.3–0.6 typically reported for the VSS and UDL formulas on similar data sets, while preserving a physically motivated description of the potential barrier for extrapolations into the unexplored superheavy region. Recent advances in machine learning have also demonstrated considerable potential for accurate predictions of \(\alpha\)-decay half-lives. Incorporating the physically motivated \(Q\)-dependent surface diffuseness as an input feature into machine-learning models may offer a promising route toward combining high predictive accuracy with improved generalizability across different nuclear regions. Such a physics-informed machine-learning framework deserves further investigation.

\section{Summary}

We have systematically investigated the influence of nuclear surface diffuseness on the interaction barriers and half-lives of 498 \(\alpha\) emitters with \(Z=61\)–118 within the spherical Coulomb and proximity potential model (CPPM), replacing the conventional fixed width \(b=1.0\) fm with \(b_{\mathrm{eff}}=(\pi/\sqrt{3})a_{\mathrm{eff}}\), where \(a_{\mathrm{eff}}=(0.0030511Z-0.001501N)\sqrt{Q_\alpha}+0.15455\) fm. In the superheavy region, the resulting diffuseness exhibits a strong correlation with \(\sqrt{Q_\alpha}\), with a global correlation coefficient of 0.968 for \(Z\geq100\) and \(b_{\mathrm{eff}}=0.69\)–0.90 fm. For the ten nuclei shared with the deformed Woods–Saxon, the prescription yields an rms deviation of just 0.030 fm and captures the variation in the benchmark diffuseness more faithfully than either the \(Z,N\)-linear or isospin-dependent parametrization. With the empirical assault frequency, the model reproduces the experimental half-lives with \(\sigma_{\mathrm{RMS}}=0.277\) and a correlation coefficient of 0.9992, with only one prediction deviating by more than an order of magnitude; the best agreement is obtained for even–even nuclei and favored \(\ell=0\) transitions. Comparisons using a common assault-frequency prescription indicate that diffuseness primarily modifies the tunneling action, whereas nucleus-dependent preformation factors account for structural effects, highlighting the need to report both assault-frequency conventions when assessing diffuseness prescriptions. These results support the independently motivated \(Q_\alpha\)-dependent parametrization as a promising choice for extrapolation to unknown superheavy nuclei. The simpler relation \(a_{\mathrm{eff}}^{(\mu)}=12.4758\mu-48.727\) provides a convenient mass-only estimate, agreeing with the \(Q_\alpha\)-dependent values to within approximately 0.03 fm for \(Z\gtrsim104\), although it is unsuitable for light nuclei. Given the spherical approximation, the empirical treatment of the assault frequency, and the limited data available at the heaviest masses, future work should incorporate deformation and parity-dependent preformation factors and extend the framework to \(^{14}\mathrm{C}\), \(^{20}\mathrm{O}\), and \(^{28}\mathrm{Si}\) cluster emission and nuclei with \(Z\geq119\), thereby providing a more systematic theoretical basis for experimental searches for the island of stability.



\end{document}